\documentclass[10pt,letterpaper]{article}
\usepackage[top=0.85in,left=2.75in,footskip=0.75in]{geometry}

\usepackage{subcaption}
\usepackage[labelformat=parens,labelsep=quad, skip=3pt]{caption}
\usepackage{graphicx}
\usepackage{verbatim}
\usepackage{amsmath,amssymb}
\usepackage{longtable}

\usepackage{changepage}
\usepackage{float}
\usepackage[flushleft]{threeparttable}
\usepackage{enumerate}

\usepackage[utf8x]{inputenc}

\usepackage{textcomp,marvosym}

\usepackage{cite}

\usepackage{nameref,hyperref}

\usepackage[right]{lineno}

\usepackage{microtype}
\DisableLigatures[f]{encoding = *, family = * }

\usepackage[table]{xcolor}

\usepackage{array}

\newcolumntype{+}{!{\vrule width 2pt}}

\newlength\savedwidth

\raggedright
\usepackage[aboveskip=1pt,labelfont=bf,labelsep=period,justification=raggedright,singlelinecheck=off]{caption}

\makeatletter
\renewcommand{\@biblabel}[1]{\quad#1.}
\makeatother

\usepackage{lastpage,fancyhdr,graphicx}
\usepackage{epstopdf}
\fancyheadoffset[L]{2.25in}
\fancyfootoffset[L]{2.25in}
\begin{document}
\vspace*{0.2in}

\begin{flushleft}
{\Large
\textbf\newline{Estimation of Ebola's spillover infection exposure in Sierra Leone based on sociodemographic and economic factors} 
}
\newline
\\

Sena Mursel\textsuperscript{1},
Nathaniel Alter\textsuperscript{2},
Lindsay Slavit\textsuperscript{3},
Anna Smith\textsuperscript{4},
Paolo Bocchini\textsuperscript{1,*}
Javier Buceta \textsuperscript{5,*},
\\
\bigskip

\textbf{1} Department of Civil and Environmental Engineering, Lehigh University, Bethlehem, PA, 18015, United States of America
\\
\textbf{2} Department of Industrial and System Engineering , Lehigh University, Bethlehem, PA, 18015, United States of America
\\
\textbf{3} Department of Chemical and Biomolecular Engineering, Lehigh University, Bethlehem, PA, 18015, United States of America
\\
\textbf{4} Department of Materials Science and Engineering, Lehigh University, Bethlehem, PA, 18015, United States of America
\\
\textbf{5} Institute for Integrative Systems Biology (I2SysBio), CSIC-UV, Paterna, VA, 46980, Spain
\\

\bigskip

%
%

* paolo.bocchini@lehigh.edu; javier.buceta@csic.es

\end{flushleft}
\section*{Abstract}
Zoonotic diseases spread through pathogens-infected animal carriers. In the case of Ebola Virus Disease (EVD), evidence supports that the main carriers are fruit bats and non-human primates. Further, EVD spread is a multi-factorial problem that depends on sociodemographic and economic (SDE) factors. Here we inquire into this phenomenon and aim at determining, quantitatively, the Ebola spillover infection exposure map and try to link it to SDE factors. To that end, we designed and conducted a survey in Sierra Leone and implement a pipeline to analyze data using regression and machine learning techniques. Our methodology is able (1) to identify the features that are best predictors of an individual's tendency to partake in behaviors that can expose them to Ebola infection, (2) to develop a predictive model about the spillover risk statistics that can be calibrated for different regions and future times, and (3) to compute a spillover exposure map for Sierra Leone. Our results and conclusions are relevant to identify the regions in Sierra Leone at risk of EVD spillover and, consequently, to design and implement policies for an effective deployment of resources (e.g., drug supplies) and other preventative measures (e.g., educational campaigns).

\section*{Introduction}
Ebola Virus Disease (EVD), more commonly referred to as Ebola, is a hemorrhagic fever pathology that causes multiorganic failure followed by death (average fatality rate $\sim 50\%$) \cite{Munoz-Fontela2017, jacob2020ebola}.
EVD originates from a virus of the {\em Filoviridae} family discovered in 1976 after two consecutive outbreaks in Central Africa \cite{chippaux2014outbreaks}.
The accumulated evidence suggest that Ebola is a zoonotic disease with main reservoir hosts being fruit bats and non-human primates \cite{fruit_bats}.
The first EVD outbreak is thought to have originated in a  cotton factory and quickly transmitted to the relatives of first patients \cite{Ebola_first_outbreak,centers_for_disease_control_and_prevention_2019}. The frequency of subsequent EVD outbreaks --approximately every other year since 1976-- as well as their locations --overwhelmingly in the sub-Saharian region-- reveals the dimension of a problem that is endemic to the African continent.
New evidence hints at the possibility of latency as one of the mechanisms to explain this endemism \cite{keita_resurgence_2021}.
As a matter of fact, at the time of preparation of this manuscript there were ongoing outbreaks in Guinea and in the Democratic Republic of Congo. Of all EVD outbreaks, the 2014--2016 one in West Africa was the most extensive and deadliest recorded ever \cite{outbreak}.
The countries most intensely hit by the outbreak were Sierra Leone, Guinea, and Liberia:
the case count of the West Africa outbreak was more than 27,000, with more than 11,000 deaths on record.
This aggravated the conditions of communities already suffering from political instability, high rates of poverty, malnutrition, low life expectancy, and weak healthcare systems \cite{omoleke2016ebola}.
The outbreak spread also outside of Africa to Europe and the USA which increased the fear of a global pandemic and resulted in extensive public and media attention; the recent COVID-19 pandemic confirms that a global outbreak in our increasingly interconnected society is a serious and realistic threat.
Indeed, the exponentially growing Ebola Virus epidemic in 2014 alarmed all the major health institutions and on August 8$^{th}$ 2014 the World Health Organization declared the EVD outbreak an international public health emergency \cite{who}.
As a result, health organizations, policy makers, and researchers were urged to understand and model the spread of Ebola in different contexts.
Modeling efforts with a predictive character aimed at mitigating the effects of the  epidemics have focused on Ebola virus pathogenicity from a molecular perspective \cite{pappalardo2017investigating, xu2020molecular}, the dynamics of the immune response \cite{martyushev2016modelling,madelain2018ebola}, human-to-human infection (including vaccination effects) \cite{desai2019real,asher2018forecasting,berge2018modeling,drake2015transmission}, the effects of human mobility \cite{li2017transmission,kraemer2019utilizing}, and also the ecology viewpoint \cite{pigott2015mapping,mombo2020detection,traditional_beliefs,fiorillo_predictive_2018}.

Interestingly, there is abundant evidence that sociodemographic and economic (SDE) factors also affect, and can be used to infer, health and health-related behaviors, including disease propagation \cite{dinh2018growth,boerma2002sociodemographic,grantz2016disparities}.
In that context, it has been shown that, typically, people with lower socioeconomic status have higher exposure to risk factors than the wealthier segments of the population \cite{feinstein1993relationship}. While a consensus on the relationship between SDE factors and exposure to infectious diseases
has not been reached \cite{grepin2020socio},
some modeling studies support the idea that poverty has an effect on the spread of infectious diseases \cite{plucinski2013clusters,bonds2012disease,plucinski2011health}. However, we point out that this relationship is mostly supported by aggregate data at the country level (e.g., GDP) and not for individuals.
Still, a number of studies have explored the correlation between disease transmission and other indicators of the individual socio-economic status \cite{iles2006socio,djiomba2019mathematical}.
In particular, Fallah et al.\ have shown, in a study based on Liberia, that individuals living in low income regions are more vulnerable to high rates of transmission and spread of Ebola \cite{fallah2015quantifying}. Moreover,
other studies concluded that the level of education is consistently associated with EVD epidemic size and spread \cite{valeri2016predicting} and that occupation is also correlated with the transmission of the Ebola virus \cite{jiang2016rapid}.

\begin{figure}[h!]
\centering
 \includegraphics[width=12cm]{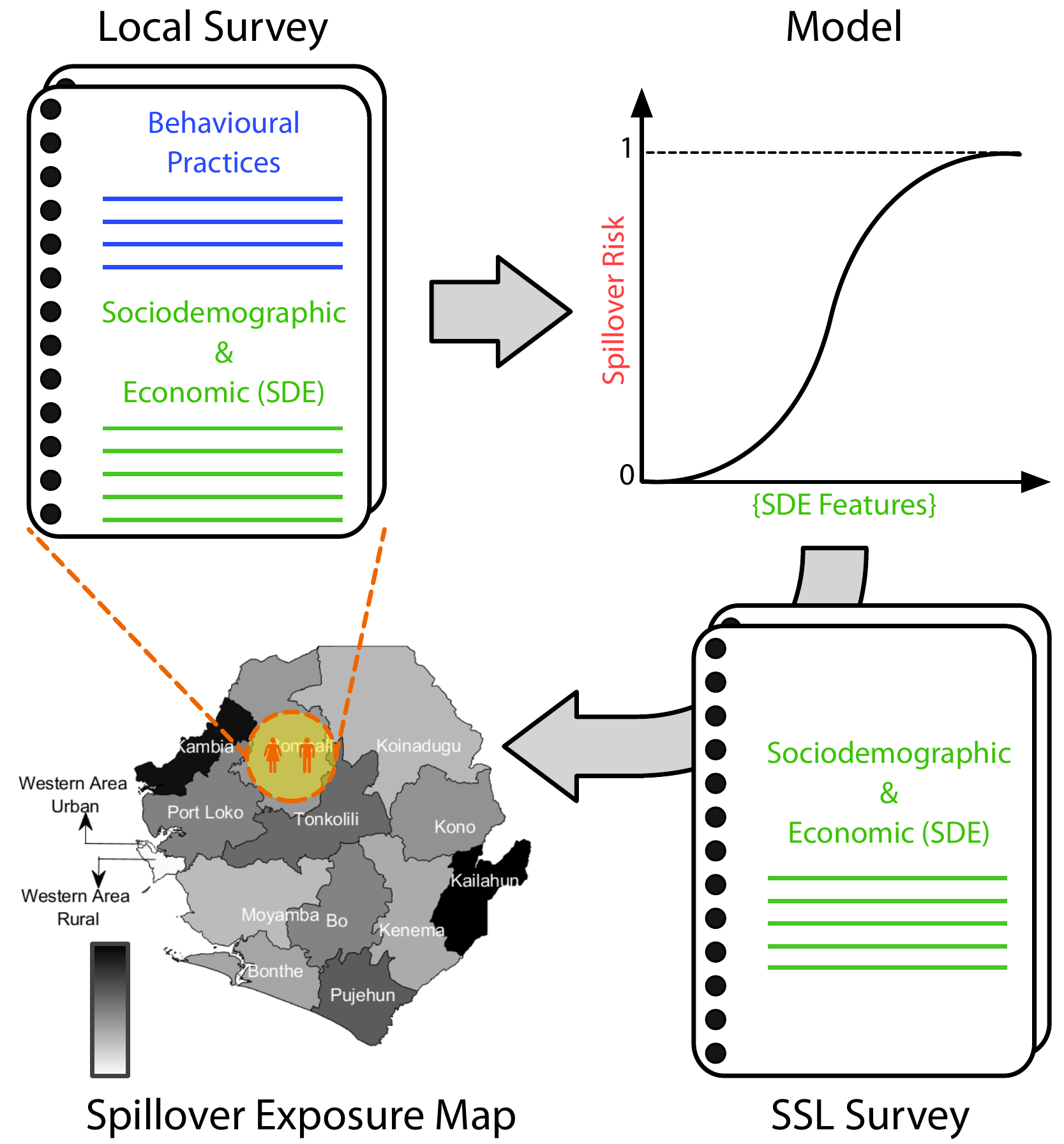}
\caption{\textbf{Methodological pipeline.} We designed a survey that combines questions about behavioural practices that could expose individuals to Ebola infection and questions to measure sociodemographic and economic (SDE) factors. The survey was administered in Sierra Leone in the Bombali rural region. We analyzed our data by different means and developed a  regression model that measures the spillover risk probability as a function of a number of SDE features. Once the model was calibrated, we extrapolated the results at the national level using surveyed data from Statistics Sierra Leone (SSL) to generate the infection spillover exposure map.}
\label{fig0}
\end{figure}

Notably, only few studies have investigated the factors contributing to the likelihood of human beings exposed to Ebola virus from animal carriers.
A recent study showed that the prominent behavioral factors associated with the transmission of the disease from animal to human (i.e., the infection spillover) are eating/hunting habits \cite{ponce2019exploring,lee-cruz_mapping_2021}. This supports previous research that indicates that direct contact with body fluids of Ebola infected animals is a substantial route of transmission \cite{osterholm2015transmission}. More recently, some surveys led to an Ebola risk score based on perceptions and knowledge about the disease.
In particular, Winters et al.\ measured the level of risk perception of survey respondents and aimed at shedding light on the relationship between risk awareness and the exposure to information sources \cite{winters2020risk}.
Also, Wille and coworkers have recently analyzed the accuracy for assessing the zoonotic risk using virological data and concluded that these analyses are incomplete and that ``surveillance at the human–animal interface may be more productive'' \cite{wille_how_2021}.

Altogether, previous works have identified determinants that increase the possibility of infection, but an association between the risky behavior of individuals and SDE factors has not been fully established. Herein we aim at bridging this gap of knowledge.
To that end, we designed, collected, and analyzed survey data from one of the regions most affected by the 2014--2016 West African Ebola epidemic.
By assessing simultaneously practices known to potentially cause
animal-to-human transmission
and socioeconomic/household traits, we define and measure, quantitatively, a spillover risk index. Since the individuals' surveyed information is regularly measured by Statistics Sierra Leone (SSL) at the nation-wide level,
our model, once calibrated, can be applied to other regions and times.
Using this approach, we extrapolated the results to the entire country of Sierra Leone, see Fig.~\ref{fig0}.
While, as reviewed above, the mechanisms driving EVD outbreaks are multifactorial, our methodology and results help to identify regions where spillovers are likely to occur. Thus, we expect our study to be relevant for EVD epidemic control, policy making, and planning to allocate resources (e.g., educational campaigns).

\section*{Methods}
\subsection*{Geographical scope of the survey}
\begin{figure}[h!]
\begin{adjustwidth}{-1.00in}{0in}
\centering

 \includegraphics[width=12cm]{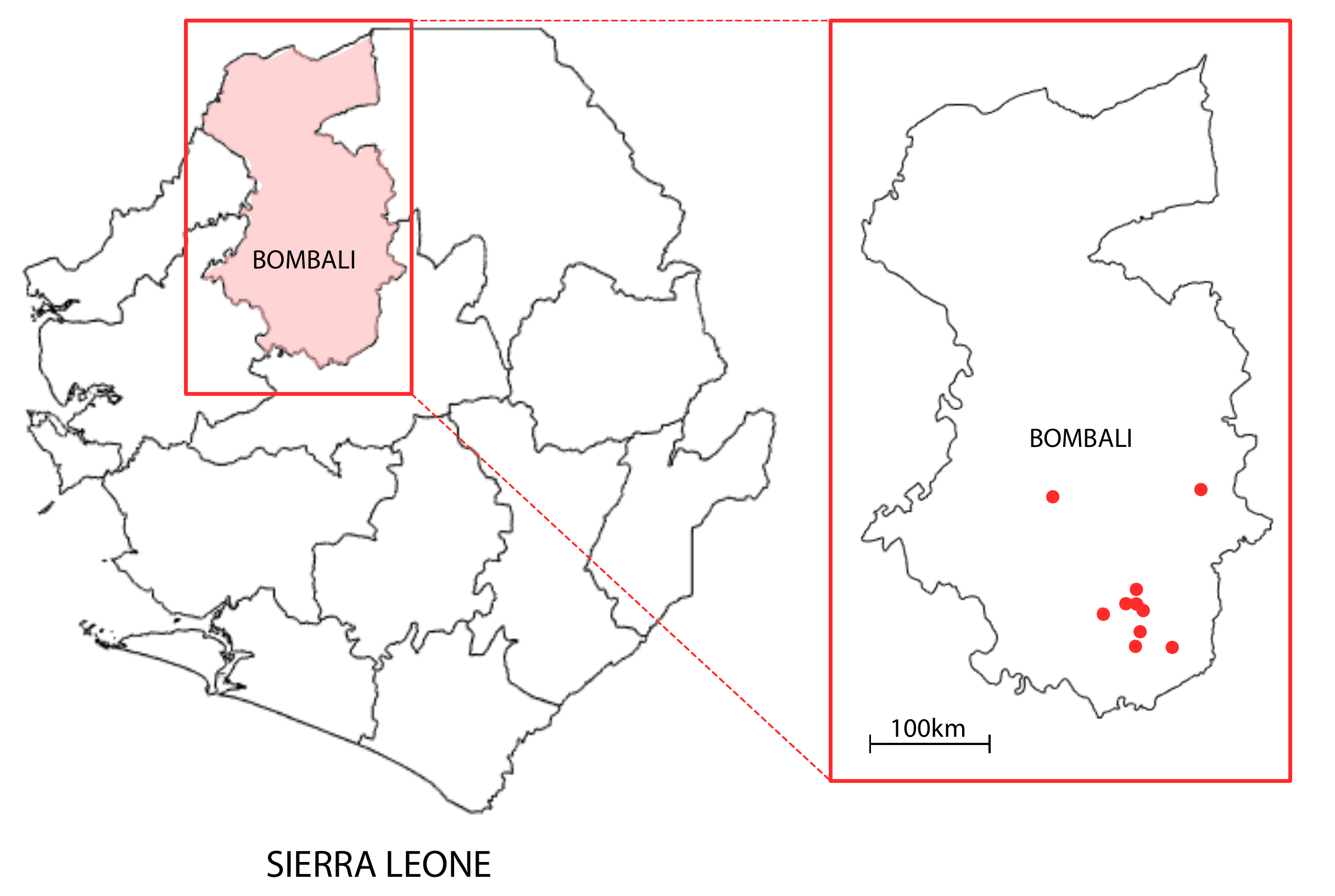}
\caption{\textbf {Survey Locations in Sierra Leone.} The survey was conducted in the district of Bombali over a period of three weeks. Ten different locations were selected (red dots) to obtain a representative sample of the population in rural areas of the country.}
\label{fig1}
\end{adjustwidth}
\end{figure}

In the summer of 2019 we carried out a survey over 3 weeks in Sierra Leone. 
Sierra Leone was selected as the country of study as it is one of the countries most severely impacted by the 2014 Ebola epidemic
\cite{costs}.
The survey was conducted in collaboration with World Hope International (WHI), a NGO
that aims at reducing poverty and improving health in Sierra Leone. 
The survey covered the district of Bombali, Fig.~\ref{fig1}. 
This district is located in the northwest region of Sierra Leone and was particularly affected by the 2014 Ebola epidemics
\cite{gleason2015establishment, centers_for_disease_control_and_prevention_2021}. 
We focused on ten different locations (a city and several villages) that were suggested by WHI authorities due to their different levels of urbanicity, most common occupation, and other demographic characteristics of the residents.
By doing this, we were able to obtain a diverse and representative sample of the population in rural areas of the country, which was our main target, due to their larger probability to have contact with wild-life (and hence  increased probability of Ebola infection due to zoonotic sources).
Over the course of the 3 weeks, 284 respondents were surveyed.
After excluding the first day respondents due to significant revisions to the survey questions (see below), 261 responses were utilized for the subsequent analyses.

\subsection*{Survey development and implementation}
The survey instrument contained five different sections: {\em i}) sociodemographic characteristics, {\em ii}) household characteristics, {\em iii}) propensity of the respondents to behavioural practices leading to some risk of Ebola infection from an animal carrier, {\em iv}) environmental characteristics, and {\em v}) perception/knowledge of EVD.
Sections {\em i}) and {\em ii}) measure SDE factors and were designed
to match the data routinely collected by Statistics Sierra Leone (SSL) as part of their Demographic and Health survey conducted once every five years. Section {\em iii}) was developed based on current knowledge about the transmission of Ebola from animal carriers to humans. 
Section {\em iv}) assessed the presence of bats and other animal carriers in the surrounding environment. Finally, section {\em v}) measured the respondent's perception and knowledge about Ebola.

In the United States, the survey was reviewed by a native from Sierra Leone to fine-tune the wording according to cultural practices and language differences (Mr.\ Vaafoulay Kanneh). 
In addition, two scholars with extensive experience on the country and its culture reviewed the questionnaire and the survey administration strategy: Prof.\ Khanjan Mehta (Vice Provost for Creative Inquiry and Director of the Global Social Impact Fellowship program at Lehigh University) and Dr.\ Soumyadipta Acharya (Graduate Program Director of the Johns Hopkins Center for Bioengineering Innovation and Design, and Instructor of Biomedical Engineering at Johns Hopkins University). 
Finally, the survey was reviewed by an independent scholar with experience in design and implementation of surveys to ensure that the questions were effectively worded and not misleading (Dr.\ Jessecae Marsh, Cognitive Psychologist and Director of the Health, Medicine and Society program at Lehigh University).

Once on the ground in Sierra Leone, WHI provided two local translators to help with the administration of the survey.
The translators were first surveyed as test subjects to confirm that the questions were clear from a Sierra Leonean perspective. 
They then translated the English version of the survey into Krio, the most commonly spoken language in Sierra Leone.
The survey was administered in the form of face-to-face interviews: the translators would ask the questions to the respondent in Krio, who would respond in Krio, and the responses were transcribed in the questionnaire by the team members from Lehigh University.

Each survey took approximately 20-30 minutes to administer. 
Team members used the application {\em Fulcrum} \cite{fulcrum}
to record the responses, register the geographical location (GPS coordinates), and the interviewee's informed consent.
Before each day of interviewing in the rural regions began, the two translators, as well as the team, would meet with the Chief of the village. 
This meeting was used to educate the leaders of the village to our presence and our purpose, as well as to get permission to conduct interviews in the village. In many cases, referencing this meeting encouraged respondents to take the survey and answer the questions more honestly.

Ethical permission for the survey (see Supplementary Material) was granted by Lehigh University's Institutional Review Board (IRB).
The project received exempt status from the IRB, and both the survey and consent statement were submitted and approved prior to the trip and after the infield changes (see below).
All survey participants were also offered paper copies of the informed consent in both English and Krio, with the contact information of the principal investigators.

\subsection*{On-site fine-tuning of the survey}
The first day of surveying took place in the city of Makeni, very close to WHI's local branch. We found that differences in African and Western cultures about the perception of ``income'' led to confusion.
We also realized that our initial strategies to test the respondents' knowledge about EVD were flawed.
For instance, asking them to list potential mechanisms of EVD contagion was leading in many case to only one item, rarely to an elaborate list.
Providing a list of actual transmission mechanisms and asking the interviewee to select if the option was correct or not led to many of them accepting systematically all options without thinking. So, to test more accurately the respondents' knowledge in the multiple-choice questions included wrong answers as possible choices. For example, in the final survey we added ``witchcraft'' as a choice of a question when we asked about possible ways of getting Ebola.
The team also found that mentioning Ebola prior to asking questions about it resulted in some discomfort that might affect the responses due to the stigma surrounding the disease throughout Western Africa. As a result of the first day of in-field experience, we decided to eliminate the surveys conducted that day (27 interviews) in future analyses and to implement some changes in the survey to alleviate the problems found. The questions regarding income were reworded to further reflect Sierra Leonean culture, the word Ebola was deliberately removed from the survey until it was specifically asked about. These changes resulted in the survey's final form (Supplementary Material) which was administered starting the second day of surveying.

\subsection*{Data preprocessing}
Survey data was a combination of quantitative and qualitative (i.e., categorical) answers as a result of the nature of our questions (see Supplementary Material).
To incorporate all qualitative answers into our quantitative model (see Results), the former were associated with binary variables as follows.
The answers to qualitative questions were grouped into categories. Then, one category, or one option in the multiple-choice questions, was chosen as the baseline. Each other option was associated with a binary variable ($1$ or $0$). As a result, the number of variables used for each question was one fewer than the number of possible categories/options, to avoid redundancy.
For example, under the work environment question, option `outdoors' was chosen as the baseline and the other option (`indoors') was associated with a binary variable. So, for this specific question, value `1' of the binary variable meant an `indoors' occupation, and value `0' meant an `outdoors' occupation.
Reference (i.e., baseline) categories/options were chosen to be either the one having largest number of responses (e.g, ``Water from a well/pump'' in the ``ways for water acquisition'' question), or the very first level of the answer options (e.g, ``no formal education'' in the ``education level'' question).

\begin{table}[b]

\centering
\caption{Water acquisition before and after data preprocessing}
\begin{tabular}{|l|l|}
\hline
\textbf{Water Acquisition Ways}  & \textbf{Assigned Categories} \\ \hline
Purchase       & water\_acquisition\_other                     \\ \hline
Running water in the house       &  water\_acquisition\_other                     \\ \hline
Water from a well/pump*          & water\_acquisition\_water\_from\_a\_well/pump                  \\ \hline
Water from a natural source
             & water\_acquisition\_water\_from\_a\_natural\_source
                     \\ \hline
\end{tabular}
 \begin{tablenotes}
     \item [] * Reference
   \end{tablenotes}
\label{Water Acquisition}
\end{table}

We set a threshold so that each possible answer category for every question has at least 10 respondents to be statistically significant. When this  criterion was not satisfied, we merged answers  into broader categories.
For example, for the ``water acquisition method'' question, only four participants declared to purchase their water, so ``Purchase" was put under the category ``water\_acquisition\_other".
Similarly, all the other options with fewer than 10 responses were assigned to the ``water\_acquisition\_other" category  (see Table ~\ref{Water Acquisition}).
For the question on the education level, as some choices had fewer than 10 responses (e.g., completed bachelors) but education levels are characterized by a clear rank, we regrouped the variables by similar levels.
For example, ``some primary school'' had fewer than 10 responses and ``completed primary school'' had more than 10 responses but, as they reveal a similar educational background, we grouped them in the same category.
We used similar approaches while categorizing the other educational options and ended up with three categories (see Table~\ref{Education}).

\begin{table}[t]
\caption{Education levels before and after preprocessing}
\begin{tabular}{|l|l|}
\hline
\textbf{Education Levels}                          & \textbf{Assigned Categories} \\ \hline
Arabic                                      & education\_primary
          \\ \hline
Completed Bachelors
                         & education\_high
   \\ \hline
Completed Diploma or Postsecondary Training
 & education\_high
  \\ \hline
Completed Junior Secondary School (JSS)
          & education\_secondary
          \\ \hline
Completed Masters or Doctorate
              & education\_high   \\ \hline
              Certificate
              & education\_high   \\ \hline
Completed Primary School                    & education\_primary
          \\ \hline
Completed Senior Secondary School (SSS)
                  & education\_secondary
         \\ \hline
Mason
                  & education\_primary
          \\ \hline
No Formal Education*                  & education\_no\_formal\_education          \\ \hline
Some primary school
                        & education\_primary
          \\ \hline
Trade school                         & education\_primary
          \\ \hline
\end{tabular}
 \begin{tablenotes}
     \item [] * Reference
   \end{tablenotes}
\label{Education}
\end{table}

For the question aiming to know the respondent's occupation, the answers were spread over 22 different options which did not reveal a clear grouping by sector. Since their combination would lead to too broad categories, which could harm the predictive capability of our model, these specific answers were ignored, and only the answers to the indoors/outdoors question (Question A7) were used to describe the occupation.
For the question asking the district of birth, 84\% of the responses were ``Bombali'', as expected. Thus, the significance of the question was deemed minimal and so we did not include it in our analysis.

Some questions implied time frequencies, such as the one about the average internet usage. In this case, the responses were converted into numerical values (between 0 and 1) that describe the number of occurrences per day, e.g., ``at least once a week'' was converted to 1/7 (see Table~\ref{Internet Use}). Numerical variables (e.g., age) were divided by their corresponding maximum values to make them dimensionless.

\begin{table}[t]
\caption{Internet use before and after preprocessing}
\begin{tabular}{|l|l|}
\hline
\textbf{Internet Use}  & \textbf{Assigned Categories} \\ \hline
At least once a day       & 1
                     \\ \hline
At least once a week          & 1/7
                  \\ \hline
At least once a month            & 1/30
                     \\ \hline
Less than once a month         & 1/60*
                       \\ \hline
Not at all               & 0
                       \\ \hline

\end{tabular}
\begin{tablenotes}
    \item [] * This was set as the average of the values in answers ``At least once  a month'' and ``Not at all''.
\end{tablenotes}
\label{Internet Use}
\end{table}

Finally, as for the location information, the GPS coordinates is available in our records and we noticed that the ``average time to highway'' responses were found to be generally inaccurate (large variability) and we checked if the responses were credible by simply measuring the distances (in miles) from households to the nearest highway. In particular, we expected that the average time to highway from similar locations (i.e., same villages) to be similar and we compared the responses with our distance measurements. We found that the coefficient of variation of the ``average time to highway'' responses located in same villages was larger than 1 in most cases. Hence, we omitted this variable (``average time to highway'') from the final data set that we used in our analyses.

In summary, taking into account the references/baselines, the final data set ended having 1 option for gender, 3 options for the education level, 1 variable for religion,  1 variable for work environment, 2 variables for relative income, 2 variables for water acquisition ways, 2 variables for ownership of cell phone as binary variables; and the frequency of internet usage, age, the number of rooms in house, the number of people in household, average time to school, average time for fuel, and average time for water as numerical variables (See Supplementary Material Dataset \#4).

\subsection*{Evaluation of the reliability of the data}
Our collected data shares sociodemographic and economic information (i.e., questions) with the surveys regularly performed by SSL (Sierra Leone Integrated Household Survey: SLIHS) \cite{odia}.
On the one hand, this allowed us to check if our survey was representative to capture the sociodemographic statistical data of the Bombali district where we ran our survey, and also of Sierra Leone in rural areas. On the other hand, as shown below, this provides the means to extrapolate the applicability of our quantitative regression model to the whole country.

\begin{figure}[h!]
\begin{adjustwidth}{-2.0in}{0in}
\centering

 \includegraphics[width=6.5in]{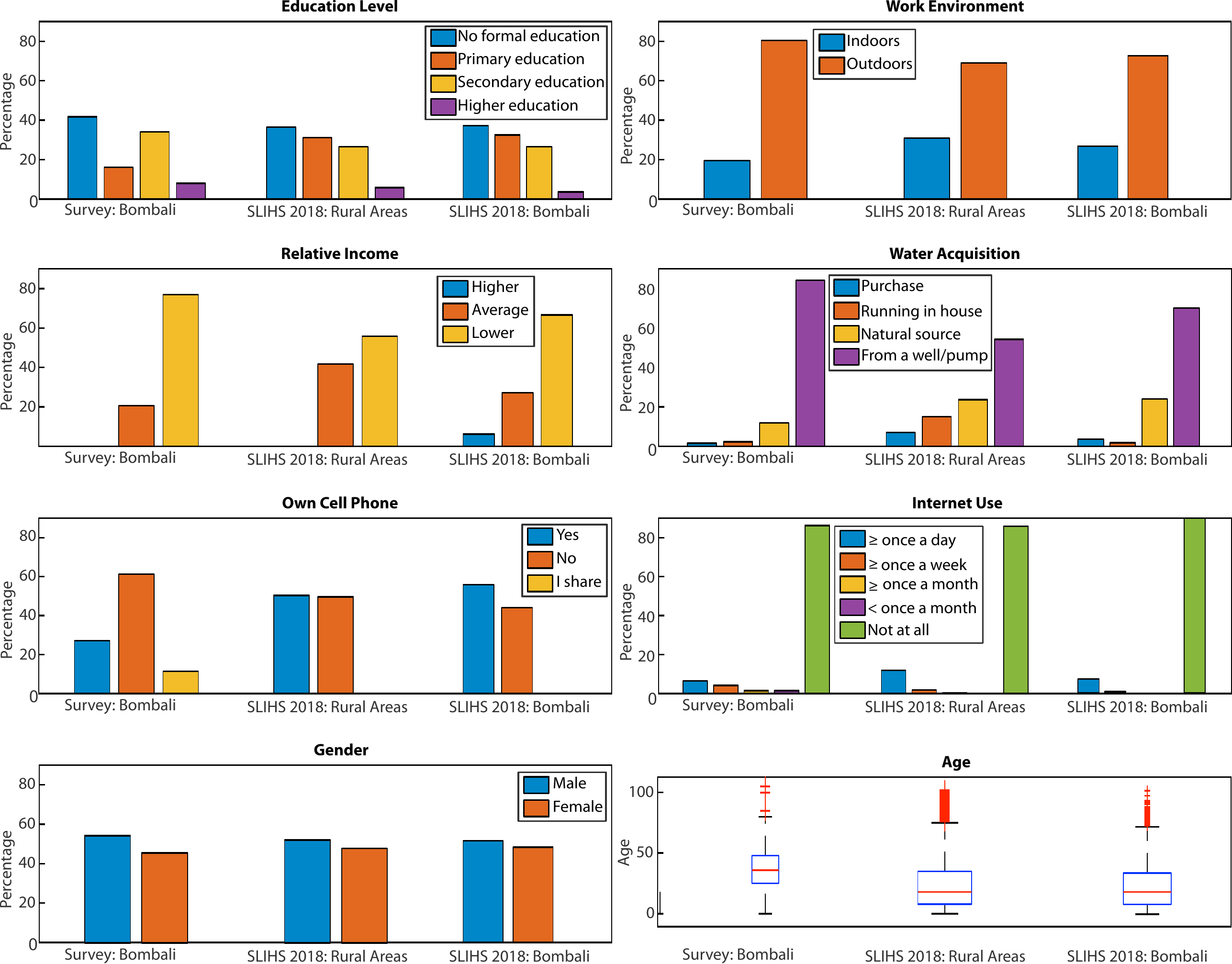}
\caption{\textbf{Comparison of the distributions in rural areas between our survey (Bombali district), SLIHS 2018 in rural areas at the country level, and SLIHS 2018 in the Bombali district.} From top to bottom and from left to right: education level, relative income, cell phone ownership, gender, work environment, water acquisition method, internet use, and boxplot of age (median: central red line; bottom and top box edges: $25^{th}$ and $75^{th}$ percentiles, respectively; outliers: plus symbols).}
\label{demographic}
\end{adjustwidth}
\end{figure}

For this comparison we used the 6 features (variables) that were deemed as representative in our regression analysis (see Results) plus ``Gender'' and ``Age'' (Figure~\ref{demographic}). Also, to compare our survey with the SLIHS 2018 at the country level, we filtered out from the data of the survey from Statistics Sierra Leone the Western Area Urban district (i.e., the capital Freetown). Including such data would mean to consider individuals with SDE characteristics that differs significantly when compared to rural areas (the focus of our research).
While there are quantitative differences, the overall trend is conserved. This suggests that our survey was representative of the Bombali district demographics and, more importantly, that our extrapolation to capture the spillover risk  at the national level is meaningful (with the exception of the Western Area Urban district that we excluded from our analysis).

\subsection*{Risk index assessment}
An important quantitative output of our survey was the Ebola spillover risk index, $R$: a number that measures the likelihood of an individual to engage in behaviors that can lead to contracting Ebola virus from an animal host.
The risk index was calculated for each individual respondent using nine questions from the section specifically related to these behaviors and five questions from the Ebola perception section. 
The contributions to the risk index resulting from these questions were assessed in different ways (see below) and provided the partial indexes $R_1$ and $R_2$ that were combined to obtain the value of $R$ for each respondent.
Table~\ref{risk_scores} collects the questions that were used to estimate $R_1$ and the scores $r_{i_{1}}$ associated with each of the possible answers: $R_1=\sum{r_{i_{1}}}$.
As shown in the table, the score for each question lies within the $[-1, 1]$ range.

{\footnotesize
\begin{center}
\begin{longtable}{|p{0.25cm}|p{5cm}|p{3cm}|p{0.75cm}|}
\caption{Risk scores $r_{i_{1}}$.} \label{risk_scores} \\

\hline \multicolumn{1}{|l|}{Question \#} & \multicolumn{1}{l|}{Question} & \multicolumn{1}{l|}{Answer} & \multicolumn{1}{|l|}{$r_{i_1}$} \\ \hline 
\endfirsthead

\multicolumn{4}{c}%
{{\bfseries \tablename\ \thetable{} -- continued from previous page}} \\
\hline \multicolumn{1}{|c|}{\textbf{Question \#}} & \multicolumn{1}{c|}{\textbf{Question}} & \multicolumn{1}{c|}{\textbf{Answer}} & \multicolumn{1}{c|}{\textbf{$r_i$}} \\ \hline 
\endhead

\hline \multicolumn{4}{|r|}{{Continued on next page}} \\ \hline
\endfoot

\hline \hline
\endlastfoot

            C2b*        & How often do you go to caves?                                                                                             & Never                                                                                       & -1.0 \\ \cline{3-4} 
            &                                                                                                                           & Every few years                                                                             & -1.0 \\ \cline{3-4} 
            &                                                                                                                           & Every few months                                                                            & 0.0 \\ \cline{3-4} 
            &                                                                                                                           & Every few weeks                                                                             & 1.0 \\ \cline{3-4} 
            &                                                                                                                           & Every few days                                                                              & 1.0 \\ \cline{3-4} 
            &                                                                                                                           & Every day/more than once per day                                                            & 1.0 \\ \cline{3-4} 
            &                                                                                                                           & Declined to answer                                                                          & 0.0   \\ \hline
C3          & How often do you wash with soap?                                                                                          & At least once a day                                                                         & -1.0        \\ \cline{3-4} 
            &                                                                                                                           & At least once a week                                                                        & 0.0  \\ \cline{3-4} 
            &                                                                                                                           & At least once a month                                                                       & 0.0  \\ \cline{3-4} 
            &                                                                                                                           & Less than a month                                                                           & 1.0  \\ \cline{3-4} 
            &                                                                                                                           & Never                                                                                       & 1.0 \\ \hline
C5          & \begin{tabular}[c]{@{}l@{}}When you eat fruit, do you check \\ if it has been bitten by animals?\end{tabular}             & Always                                                                                      & -1.0 \\ \cline{3-4} 
            &                                                                                                                           & Sometimes                                                                                   & 0.0 \\ \cline{3-4} 
            &                                                                                                                           & Never                                                                                       & 1.0 \\ \cline{3-4} 
            &                                                                                                                           & I don't eat fruit                                                                           & 0.0 \\ \hline

C9          & How often do you eat bushmeat?                                                                                            & Every Meal                                                                                  & 1.0 \\ \cline{3-4} 
            &                                                                                                                           & Once a day                                                                                  & 1.0 \\ \cline{3-4} 
            &                                                                                                                           & At least once a week                                                                        & 1.0 \\ \cline{3-4} 
            &                                                                                                                           & At least once a month                                                                       & 1.0 \\ \cline{3-4} 
            &                                                                                                                           & At least once per year                                                                      & 0.5 \\ \cline{3-4} 
            &                                                                                                                           & Never                                                                                       & -1.0 \\
            &                                                                                                                           & I used to, but no longer                                                                    & -1.0 \\ \hline
C11         & \begin{tabular}[c]{@{}l@{}}Do you clean your hands before \\ eating?\end{tabular}                                         & Always                                                                                      & -1.0 \\ \cline{3-4} 
            &                                                                                                                           & Sometimes                                                                                   & 0.0 \\ \cline{3-4} 
            &                                                                                                                           & Never                                                                                       & 1.0 \\ \hline

C13         & \begin{tabular}[c]{@{}l@{}}How often do you spend time \\ in places where bats nest?\end{tabular}                         & Never                                                                                       & -1.0 \\ \cline{3-4} 
            &                                                                                                                           & Every few years                                                                             & 0.0 \\ \cline{3-4} 
            &                                                                                                                           & Every few months                                                                            & 0.5 \\ \cline{3-4} 
            &                                                                                                                           & Every few weeks                                                                             & 1.0 \\ \cline{3-4} 
            &                                                                                                                           & Every few days                                                                              & 1.0 \\ \cline{3-4} 
            &                                                                                                                           & Every day/more than once per day                                                            & 1.0 \\ \hline
            C14         & \begin{tabular}[c]{@{}l@{}}How often do you have contact\\ with someone else’s blood \\ or bodily fluids?\end{tabular}      & At least once a day                                                                         & 1.0 \\ \cline{3-4} 
            &                                                                                                                           & At least once a week                                                                        & 1.0 \\ \cline{3-4} 
            &                                                                                                                           & At least once a month                                                                       & 0.0 \\ \cline{3-4} 
            &                                                                                                                           & Less than a month                                                                           & 0.0 \\ \cline{3-4} 
            &                                                                                                                           & Never                                                                                       & -1.0 \\ \hline
C15         & \begin{tabular}[c]{@{}l@{}}Do you believe that touching\\ raw meat or any live animal \\ could spread disease?\end{tabular} & Yes                                                                                         & -1.0 \\ \cline{3-4} 
            &                                                                                                                           & No                                                                                          & 1.0 \\ \cline{3-3}
            &                                                                                                                           & I don't know                                                                                & 0.5 \\ \hline
C16         & \begin{tabular}[c]{@{}l@{}}Do you believe that eating \\ bushmeat could spread disease?\end{tabular}                      & Yes                                                                                         & -1.0 \\ \cline{3-3}
            &                                                                                                                           & No                                                                                          & 1.0 \\ \cline{3-3}
            &                                                                                                                           & I don't know                                                                                & 0.0 \\ \hline

E2          & \begin{tabular}[c]{@{}l@{}}Do you think a person could \\ get Ebola from an animal, dead \\ or alive?\end{tabular}            & Yes                                                                                         & -1.0 \\ \cline{3-4} 
            &                                                                                                                           & No                                                                                          & 1.0 \\ \cline{3-4} 
            &                                                                                                                           & I don't know                                                                                & 0.5  \\ \hline
           
E7          & \begin{tabular}[c]{@{}l@{}}Do you believe that you \\ can get Ebola from bushmeat?\end{tabular}                           & Yes                                                                                         & -1.0 \\ \cline{3-4} 
            &                                                                                                                           & No                                                                                          & 1 \\ \cline{3-4} 
            &                                                                                                                           & I don't know                                                                                & 0.5 \\ \hline
\end{longtable}
\begin{tablenotes}
     * This question was asked only to the participants who answered ``Yes'' to  question C2, ``Do you know any caves?''.
   \end{tablenotes}
\end{center}
}

Specifically, every answer gets a score of $-1$, $-0.5$, $0$, $0.5$ or $1$ depending on the level of exposure to infection: if an action reveals a risky behavior, we assigned a score of $1$ and if the behavior decreases the likelihood of infection, then $-1$ was assigned.
For questions where answers imply a time frequency (e.g., ``every day''), the score of the riskiest answer was given $1$ and the score of $-1$ was assigned to the least risky answer (intermediate answers were given one of the other five possible values mentioned above).
The second contribution to the risk index, $R_2=\sum{r_{i_{2}}}$, was determined based on ``check all that apply'' type of questions (questions E1, E2b, and E3 of the survey, Table~\ref{risk_scores_2}).
The possible options for these three questions included both correct and wrong answers  on mechanisms of human-to-human Ebola infection, animal-to-human Ebola infection, and strategies to prevent Ebola. 
As mentioned above, wrong answers were included in these questions after we evaluated the conducted interviews of the first day and we noticed that a number of respondents checked all choices.
We then modified the questions by providing multiple options that included both correct and wrong answers. 
Using the modified survey, the scores $r_{i_2}$ were assigned using the following procedure. 

\begin{enumerate}[i.]
    \item If a respondent gave more than one wrong answers to a question, then $r_{i_2}=1$.
    \item If a respondent gave only one wrong answer to a question and could not provide at least half of the reasonable answers, then $r_{i_2}=1$.
    \item If a respondent gave only one wrong answer to a question but provided at least half of the reasonable answers, then $r_{i_2}=0.5$.
    \item If a respondent gave only correct answers, then $r_{i_2}=-1$.
    \item If a respondent answered ``I don't know'', then $r_{i_2}=0.5$.
\end{enumerate}

{\footnotesize
\begin{table}[]
\caption{Risk scores $r_{i_{2}}$.} \label{risk_scores_2}
\begin{adjustwidth}{-1.0in}{}
\begin{tabular}{|l|l|l|}
\hline
\cellcolor[HTML]{FFFFFF}{\color[HTML]{000000} Question \#} & Question                                                                                                                               & Answer                                        \\ \hline
E1                                                         & \begin{tabular}[c]{@{}l@{}}What are the ways in which a \\ person gets Ebola?(Check all that \\ apply) (Open Question)\end{tabular}    & By air                                        \\ \hline
                                                           &                                                                                                                                        & Bad odor or smell                             \\ \cline{3-3} 
                                                           &                                                                                                                                        & Preparing bushmeat as a meal                  \\ \cline{3-3} 
                                                           &                                                                                                                                        & Eating bushmeat                               \\ \cline{3-3} 
                                                           &                                                                                                                                        & Eating fruits likely to have bitten by bats   \\ \cline{3-3} 
                                                           &                                                                                                                                        & Eating with an infected person                \\ \cline{3-3} 
                                                           &                                                                                                                                        & The saliva of an infected person              \\ \cline{3-3} 
                                                           &                                                                                                                                        & Blood of an infected person                   \\ \cline{3-3} 
                                                           &                                                                                                                                        & The sweat of an infected person               \\ \cline{3-3} 
                                                           &                                                                                                                                        & The urine of an infected person               \\ \cline{3-3} 
                                                           &                                                                                                                                        & Feces of an infected person                   \\ \cline{3-3} 
                                                           &                                                                                                                                        & Living with an infected person                \\ \cline{3-3} 
                                                           &                                                                                                                                        & Working with an infected person               \\ \cline{3-3} 
                                                           &                                                                                                                                        & God's will                                    \\ \cline{3-3} 
                                                           &                                                                                                                                        & Witchcraft                                    \\ \cline{3-3} 
                                                           &                                                                                                                                        & Government hoax                               \\ \cline{3-3} 
                                                           &                                                                                                                                        & Ebola does  not exist                         \\ \cline{3-3} 
                                                           &                                                                                                                                        & I do not know                                 \\ \cline{3-3} 
                                                           &                                                                                                                                        & Declined to answer                            \\ \hline
E2b                                                        & \begin{tabular}[c]{@{}l@{}}How could a person get Ebola \\ from an animal? (Check all that \\ apply) (Read options)\end{tabular}       & Having an animal as a pet                     \\ \hline
                                                           &                                                                                                                                        & Eating any meat                               \\ \cline{3-3} 
                                                           &                                                                                                                                        & Eating bushmeat                               \\ \cline{3-3} 
                                                           &                                                                                                                                        & Watching an animal                            \\ \cline{3-3} 
                                                           &                                                                                                                                        & Eating fruits bitten by an animal             \\ \cline{3-3} 
                                                           &                                                                                                                                        & Hunting                                       \\ \cline{3-3} 
                                                           &                                                                                                                                        & Preparing bushmeat as a meal                  \\ \hline
E3                                                         & \begin{tabular}[c]{@{}l@{}}In general, how do you think a\\ person avoids Ebola? (Check all \\ that apply) (Read Options)\end{tabular} & Brushing their teeth                          \\ \hline
                                                           &                                                                                                                                        & Sleeping under a mosquito net                 \\ \cline{3-3} 
                                                           &                                                                                                                                        & Avoiding contact with blood and bodily fluids \\ \cline{3-3} 
                                                           &                                                                                                                                        & Drinking tea                                  \\ \cline{3-3} 
                                                           &                                                                                                                                        & Staying inside when it rains                  \\ \cline{3-3} 
                                                           &                                                                                                                                        & Not touching anyone with the disease          \\ \cline{3-3} 
                                                           &                                                                                                                                        & Cleaning themselves with soap and water       \\ \cline{3-3} 
                                                           &                                                                                                                                        & Avoiding funerals or burial rituals           \\ \cline{3-3} 
                                                           &                                                                                                                                        & Drinking only tap water                       \\ \cline{3-3} 
                                                           &                                                                                                                                        & Avoiding the forest/woods                     \\ \cline{3-3} 
                                                           &                                                                                                                                        & I don’t know                                  \\ \cline{3-3} 
                                                           &                                                                                                                                        & Declined to answer                            \\ \hline
\end{tabular}
\end{adjustwidth}
\end{table}
}

\subsection*{Regression analysis}
One goal of our study was to develop a methodology able to determine the risk index $R$ not just for individuals that took our survey but also for individuals for which SDE information is part of the publicly available data from Statistics Sierra Leone (SSL).
To that end, we calibrated a model that takes as \textit{input} the answers to the same SDE questions from the survey of SSL and returns as \textit{output} the risk index, $R$.
We calibrated and tested multiple models via regression analysis and supervised machine learning, where the risk index was used as a response variable for training and the other answers were used as features.
We notice that when using these approaches, the models were not trained using the actual value of the risk index $R$.
Instead of the continuous description of the risk index, we splitted the respondents into ``high risk of spillover exposure'' and ``low risk of spillover exposure'', based on whether $R$ was above or below the average risk index.
In this way, we simplified the output of the predictive algorithm, settling for a classification (high/low risk), rather than a full quantification of $R$.
However,  our analyses in that regard showed little to minimum accuracy.
After extensive testing and refinement, the best result was achieved with the extensive gradient boosted decision trees, which led to a root mean squared of error (RMSE) of 0.47 on the testing data and was deemed unacceptable as the output is given as binary.
Our conclusion was that all machine learning approaches were inconclusive, arguably due to the small number of observations available for training.
Additionally, data visualization techniques, including principal component analysis (PCA) and uniform manifold approximation and projection (UMAP) to reduce  dimensionality were applied. 
However, no distinct clusters were observed (see Supplementary Material).

For this reason, we focused on regression models, which tend to have fewer parameters to calibrate, and provide workable results even with small training datasets.
For this task, we used the R statistical programming environment \cite{R}. 
Our initial attempt was to use a multiple linear regression.
A total of 19 input variables was available in the post-processed survey data  and  we used the \textit{regsubsets} function of the contributed R package \textit{leaps} which performs an exhaustive search for the best subsets of the variables in the dataset to predict the risk index. The best model was found to have only 5 variables.
The model performance was evaluated by computing the adjusted $R^2$ value: the best model resulted in an adjusted $R^2$ of $0.073$, which was clearly too low.
Hence, we tried to calibrate a model with logistic regression.

For the logistic regression, the value of the risk score, $R$, had to be converted into a dichotomous variable that describes if respondent either does or does not engage in behaviors that leads to risk of Ebola infection. Thus, we first scaled and normalized $R$ with respect to its minimum and maximum values:

\begin{equation*}
    R_n=\dfrac{R-\min(R)}{\max(R)-\min(R)}
\end{equation*}

Second, by using this normalized value of the risk index, $R_n\in\left(0,1\right)$, we set a cutoff value of 0.5 that allowed to classify individuals in a binary way: individuals that engaged in a risky behavior ($R_n > 0.5$, high risk) and individuals that did not engage in a risk behavior ($R_n < 0.5$, low risk) from the viewpoint of a possible Ebola infection.

In our logistic regression model, the outcome variable, $Y$, is described as,

\begin{equation}
Y=logit(R_n) = \log \left(\dfrac{R_n}{1-R_n}\right)=\beta_0+\beta_1X_1+\beta_2X_2+ \ldots +\beta_n X_n
\label{logistic_regression_equation}
\end{equation}

As shown in the Results section, this regression model provided satisfactory predictive capabilities.

\section*{Results}
\subsection*{Sociodemographic and economic factors underlying the Ebola spillover risk}
Following a classification of the spillover risk index into a binary class (high/low risk), we were able to implement a logistic regression (Methods) and investigated its predictive accuracy and the optimal subset of features to be included.
The feature subset was found based on the Akaike Information Criterion (AIC), that estimates the prediction error: 
the model giving the smallest AIC value was selected \cite{akaike1974new}. 
Forward and backward stepwise logistic regression through AIC
were applied to select the optimum number of independent variables and to eliminate the variables not contributing significantly to the exposure to risk of spillover.

\begin{figure}[h!]
\begin{adjustwidth}{-2.00in}{0in}
\centering

 \includegraphics[height=16.5cm]{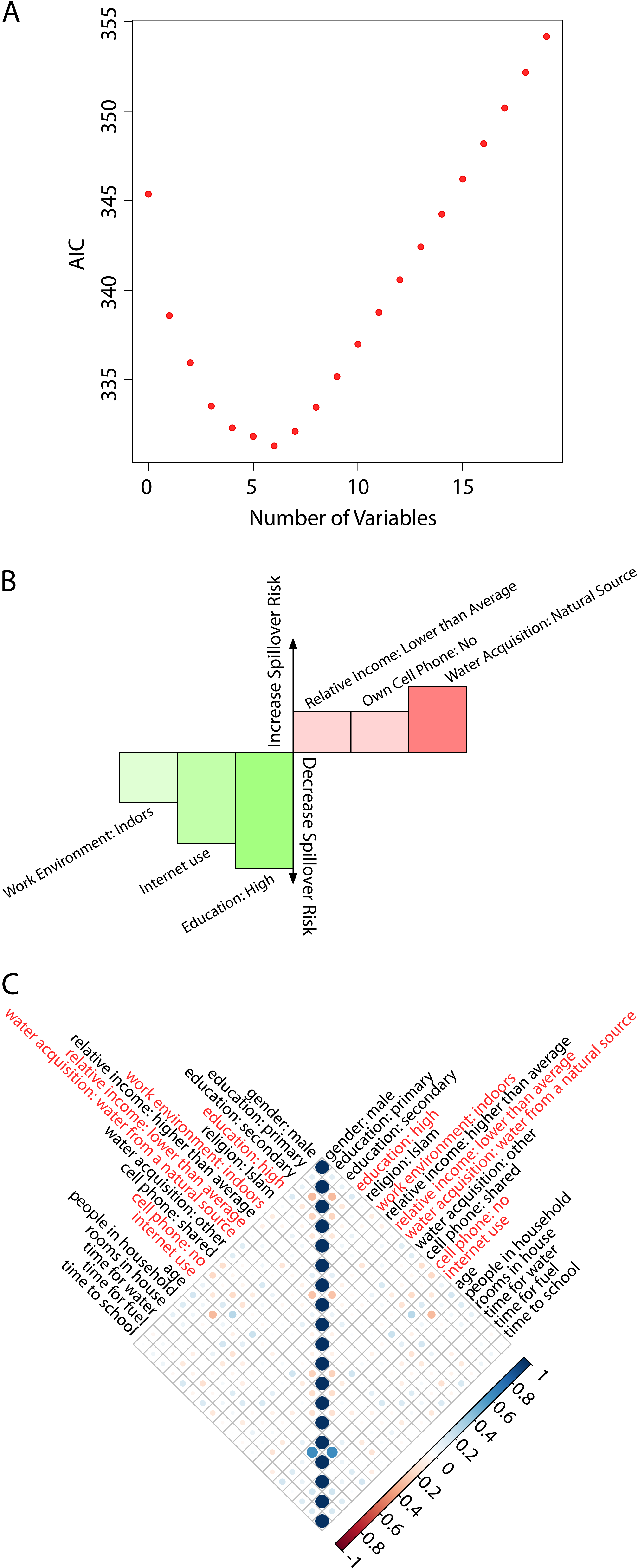}
 \end{adjustwidth}
\caption{\textbf {A: AIC values as a function of the number of variables (features).} Either starting from a null model and increasing the number of features (forward stepwise logistic regression) or from a complete model and decreasing the number of features (backward stepwise logistic regression), we consistently found that a model with six variables shows a global minimum for AIC (minimum prediction error). \textbf{B: Graphical representation of the logistic regression coefficients.} Magnitude of the $\beta_{i}$ coefficients (normalized to the maximum) and their sign (positive/negative: red/green). The selected features balance SDE factors that increase or decrease the spillover risk. \textbf{C: Graphical representation of the correlation matrix among variables.} Our analysis indicates that there is no significant correlation among variables (red text stand for the selected features in the logistic regression).}
\label{AIC}

\end{figure}

Our analyses concluded that a model with six (out of nineteen) features provided a global minimum for the AIC value (Figure \ref{AIC}A and Table \ref{logistic}).
Since the adjusted $R^2$ cannot be used as indicator of the goodness of fit using a logistic regression, we used instead the model accuracy, defined as the percentage of cases where the binary output variable (high/low risk) is correctly predicted by the model.
We point out that to measure the model accuracy and robustness we performed a 10-fold cross-validation that we repeated three times with different data partitioning, for a total of $30$ analyses using $10\%$ of the data as test samples each time.
The accuracy level ranged from $0.5$ to $0.81$, with an average accuracy of $0.657\pm 0.07$.
Based on these results, we concluded that the model is accurate and  robust.

As shown in Table~\ref{logistic} and Figure~\ref{AIC}B, the best indicators of SDE factors able to capture the Ebola spillover risk are features related with education level, work environment, income (including measures of purchasing power), and  access to information.

\begin{center}
\begin{table}[!ht]

\begin{adjustwidth}{-1.0 in}{0in} 
\centering
\caption{
{\bf Selected SDE features with the best predictive capabilities in the logistic regression model.}}
\begin{tabular}{|l|l|l|l|l|l|l|}
\hline
{\bf Feature} & {\bf $\beta_i$}& {\bf $p$-value}\\ \hline
education: high & $-1.4\pm 0.8$ &  $0.07239$\\ \hline
work environment: indoors & $-0.6\pm 0.4$ &  $0.09749$\\ \hline
internet use & $-1.1\pm 0.8$ &  $0.1763$\\ \hline
relative income: lower than average & $0.5\pm 0.3$ &  $0.1185$ \\ \hline
water acquisition: natural source & $0.8\pm 0.4$ & $0.04261$\\ \hline
own cell phone: no & $0.5\pm 0.3$ &  $0.09444$\\ \hline
\end{tabular}
   
\caption*{Value of the coefficients $\beta_i$ for the logistic model shown in Eq.~\eqref{logistic_regression_equation}. $\pm$ ranges show the standard error of the corresponding coefficients.}

\label{logistic}
\end{adjustwidth}
\end{table}
\end{center}

The sign of the coefficient associated with each feature is indicative of the feature being associated with high (positive sign) or low (negative sign) the Ebola spillover risk. In that regard, our results revealed that work conditions that decrease possible contact with animals, better education background, and access to information are factors that decrease the spillover risk. On the other hand, a worse economic status and activities that imply contact with the natural environment increase the chances of infection from a zoonotic source (Fig. \ref{AIC}B).
To investigate the possible interdependence among predictor variables, we computed their correlation matrix (Figure \ref{AIC}B). No strong correlation between any pairs was found and the more significant ones are consistent with our expectations (e.g., highest correlation coefficient: $0.63$ between ``people in household'' and ``rooms in house'').

\begin{figure}[h!]
\begin{adjustwidth}{-1.00in}{0in}
\centering

 \includegraphics[width=15cm]{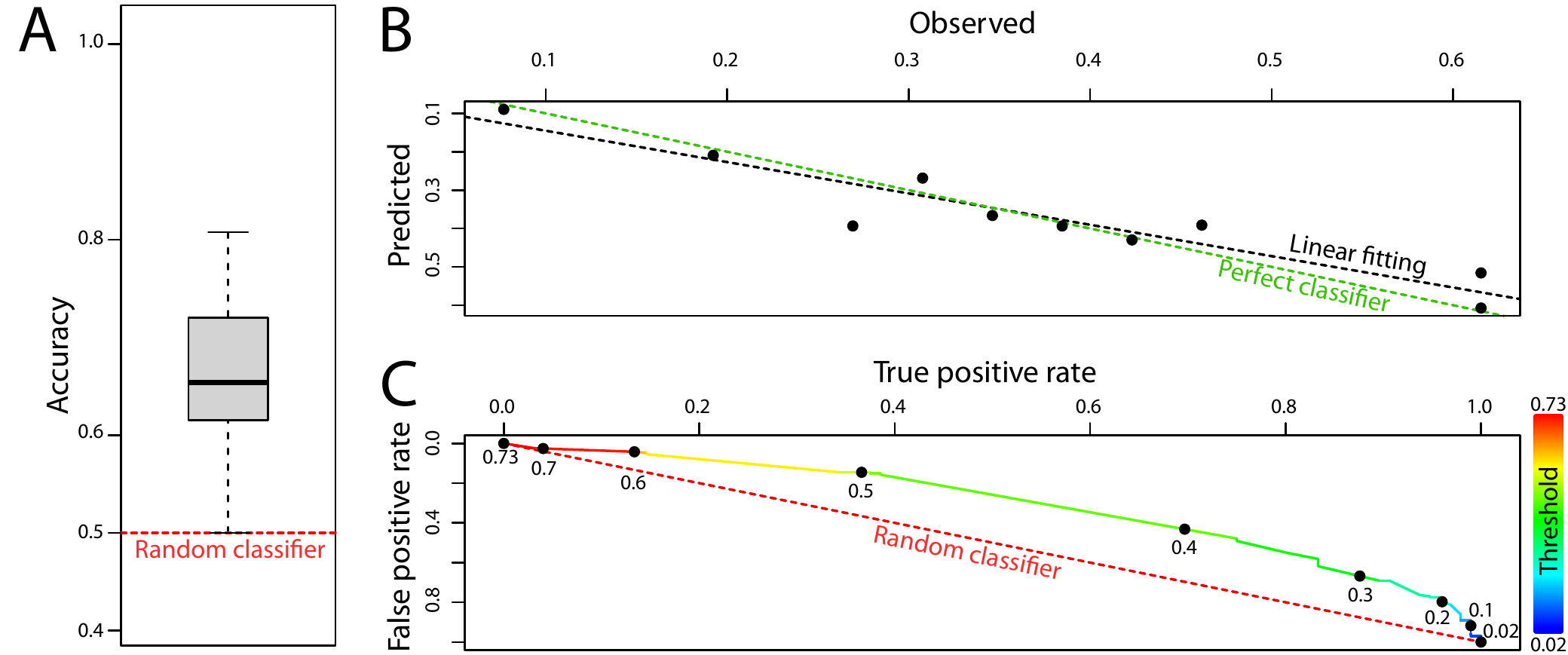}
\end{adjustwidth}
\caption{\textbf {A: Box plot of the accuracy of the logistic model.} The accuracy, measured as the fraction of correctly predicted spillover risk, is $0.657\pm 0.07$. In the plot the wide black line indicates the median. The box delimits the $\left(25\%,75\%\right)$  percentile interval, and the whiskers represent the minimum and maximum values (no outliers were present in this case). The accuracy analysis was performed repeating a 10-fold cross validation three times (see text). \textbf{B: Predicted versus observed spillover risk scores.} The green dotted line is the expected behavior of a perfect classifier and the circles represent the results obtained from our model (see text). The black dotted line is the linear fitting of the points. \textbf{C: ROC curve.} As a function of the classifier threshold (color scale) the true versus false positive rate is plotted. The model deviates clearly from a random classifier (red dotted line). Analyses with a threshold larger/smaller than $0.73$/$0.02$ accumulate in top left/bottom right corner of the plot.}
\label{model results}
\end{figure}

We further tested the validity of our logistic model in terms of its predictive capability by different means. To assess the goodness of fit we used the Hosmer-Lemeshow test \cite{archer2006goodness} that calculates the discrepancy between the predicted and observed risk indexes. The result from the test was not significant ($\chi^2=2.8848$) and indicated a satisfactory predictive power ($p=0.9414>0.05$). The successful calibration of predictions was confirmed by analyzing the predicted versus  observed risk score (Figure \ref{model results}B).
To that end, we ordered the interviewees by their predicted spillover risk and divided the sorted data into ten equal sets (deciles or bins). For each of these sets we compared the predicted versus  observed spillover risk. This analysis confirmed that the regression model is reliable (Fig. \ref{model results}B). Also, given that our model aims at discriminating between the values of a binary outcome (i.e., high risk or low risk), we computed the Receiver Operating Characteristic (ROC) curve \cite{ROC_curve} in Fig.~\ref{model results}C. Our model deviates from a random classifier in a satisfactory way and the restult of this analysis contributes to justifying the value of the threshold used in the logistic classification (i.e., $0.5$). As a way to measure the goodness of the predictive character of our model, we computed the area under the ROC curve (AUC): a perfect classifier would give a value of $1$ for this measure and a random classifier a value of $0.5$. 
In our case we obtained $0.69$, which was considered acceptable.

In summary, our logistic regression model is able to identify a reduced set of SDE features to quantify with enough accuracy and in a robust way the Ebola spillover risk in individuals. As shown below, this calibrated model was subsequently used to extrapolate the analysis to the entire country.

\subsection*{Application of the risk model nation-wide: infection spillover exposure map}

Once our predictive model was properly calibrated and deemed reliable, we aimed at applying it to the entire nation of Sierra Leone. To that end, we used data from the broader survey (SLIHS) conducted by SSL in 2018 and for which responses of individuals are publicly available ($\sim 4\cdot 10^{4}$ interviewees). 
We designed our survey to include some of the SDE questions in the SLIHS survey. 
Consequently, we were able to use the SLIHS data set as input in our model and estimate the risk scores of each respondent. 
As for how representative was our data set compared to the overall statistics to justify this extrapolation, our analyses indicated that we captured sufficiently well the demographics of rural areas of Sierra Leone (see Methods).

We performed our calculations at the district, $d$, level by computing for each individual, $i$, the spillover risk index using our logistic model: $R_{n}|_{i,d}$. By setting a threshold of $0.5$ (Methods, see also Fig. \ref{model results}C), the fraction  of surveyed individuals at risk of infection in a district reads:
\begin{equation}
p_{d}=\frac{1}{N_{d}}\sum_{i=1}^{N_{d}}{\theta\left(R_{n}|_{i,d}-0.5\right)},
\end{equation}
where $\theta\left(\cdot\right)$ is the Heaviside step function and the sum runs over the $N_d$ individuals that were surveyed in the district. Thus, the density of individuals at risk of being exposed to spillover infection in a district,$\rho^{I}_{d}$, is
\begin{equation}
\rho^{I}_{d}=p_{d}\rho_{d},
\end{equation}
$\rho_{d}$ being the population density in the district \cite{census}. Thus, the infection spillover exposure map is, effectively, the population density map modulated by the spillover risk probability.

\begin{figure}[h!]
\centering
\begin{adjustwidth}{-1.00in}{0in}

 \includegraphics[width=14cm]{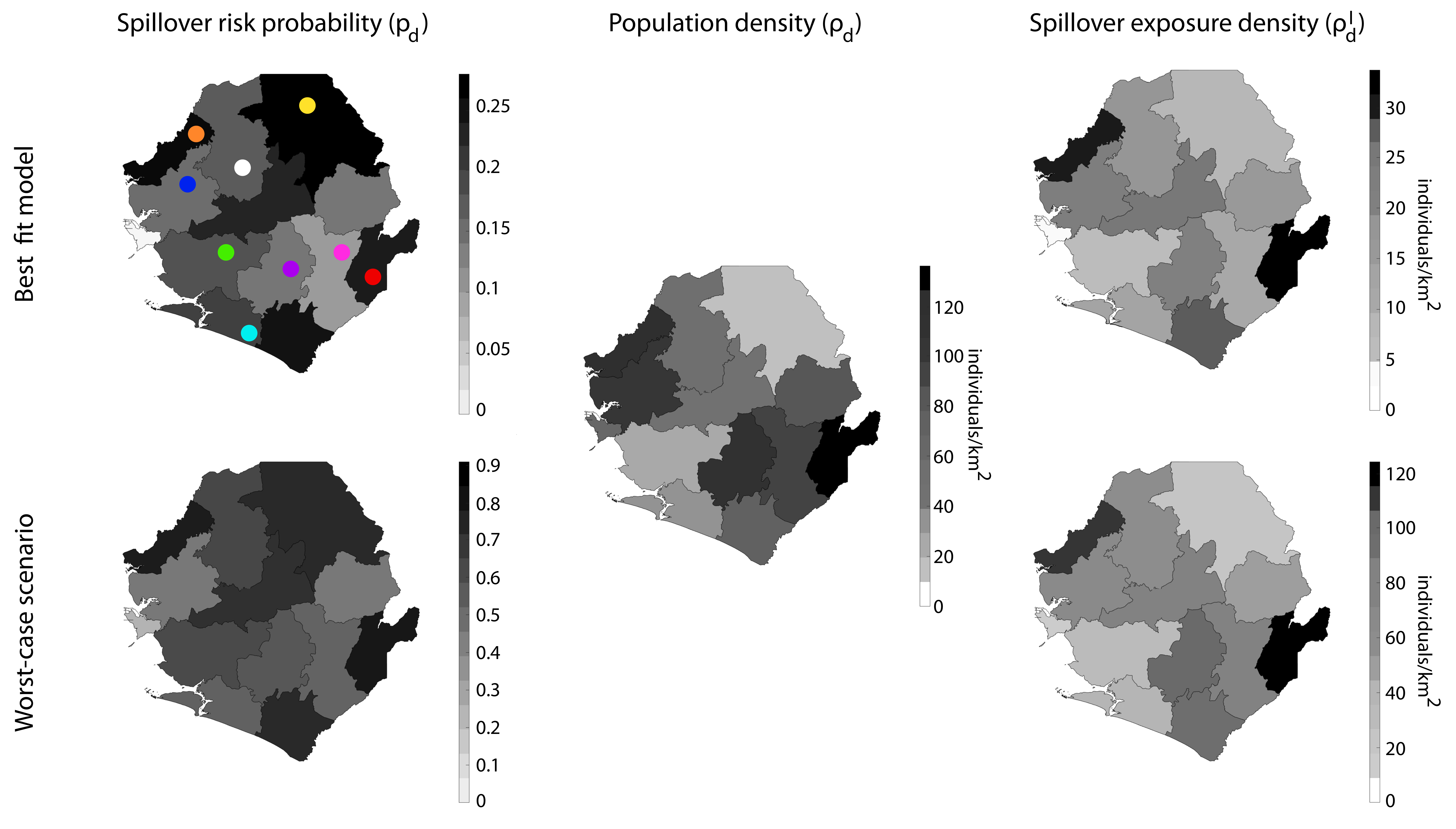}
 \end{adjustwidth}
  \caption{\textbf{Estimation of the infection spillover map in Sierra Leone by districts.} From left to right the figure shows the spillover risk probability ($p_d$), the population density ($\rho_d$), and the infection spillover exposure ($\rho_{d}^{I}$) respectively. In the case of $p_d$ and $\rho_d$ the maps showed on the top stand for the cases of the best fit logistic model and on the bottom the worst-case scenario (see text). District color codes (as shown on top left): Bo (purple), Bombali (white), Bonthe (cyan), Kailahum (red), Kambia (orange), Kenema (pink), Koinadugu (yellow), Moyanba (green), and Port Loko (blue).}
\label{map_prob}
\end{figure}

Figure \ref{map_prob} shows
the infection spillover exposure maps, $\rho_{d}^{I}$, by taking into account the values of $\beta_i$ in the logistic regression (Table \ref{logistic}) and also the worst-case scenario. To compute the worst-case scenario we used as model coefficients the values  $\beta_i + \epsilon_i$ ($\epsilon_i$ being the error of the coefficient $\beta_i$).
We point out that the best-case scenario computed by using $\beta_i - \epsilon_i$ predicts no spillover infection, so the associated maps are not included (see Discussion).

Our data and analyses suggest that Kailahun and Kambia are the rural districts in Sierra Leone with the highest density of individuals exposed to an infection spillover due to SDE factors. This is a combined effect of both, high risk spillover probabilities and high population densities. Kailahun is in fact the district where the 2014 Ebola epidemics originated  \cite{fang2016transmission}. Koinadugu and Moyamba are two of the districts with a spillover risk probability that is significantly large. However, their low population density contributes to decrease their spillover exposure. A similar behavior was observed in Bonthe district. However, in Port Loko and Bo districts the opposite behavior was found: not excessively large risk probabilities but high population densities modulate each other and contribute to leave the spillover exposure at average levels. 
The district of Kenema, that was one of the most severely affected by the 2014 epidemics \cite{nanyonga2016sequelae}, is not revealed as one of the districts with higher exposure. We notice however that our model does not account for human-human infective processes and, consequently, this result is not particularly surprising. Still, we point out that Kenema neighbors Kailahun that, as mentioned above, is one of the districts with a larger spillover exposure and, arguably, the combined effect of spillover exposure due to zoonotic sources with mobility and human-human infection would have contributed in the past to the large levels of EVD in Kenema. As for the district of Bombali where we ran our survey, average risk probability and population density lead to average spillover risk. Finally, we did not observed significant qualitative changes in the spillover risk probability between the best model and the worst-case scenarios. Nonetheless, we stress the noticeable large levels of spillover risk probability in many district of the country even in the best model scenario. This points out the necessity, according to our study, of implementing measures that could contribute to lower the spillover risk probability (see Discussion).

\section*{Discussion and conclusions}
Herein we have proposed for the first time, to be best of our knowledge, a methodological pipeline to quantify the infection spillover risk probability in individuals and the spillover exposure map at the country level due to SDE factors. Our research contributes to the recent interest in understanding the complexity of epidemic propagation due confluent effects and where SDE factors have been proved to be relevant and yet often disregarded. In that regard, previous approaches have focused on evaluating and weighting these factors globally (e.g., at the country level). We instead have focused on the individual level. The advantage of our approach is that it allows scholars and decision makers to obtain a deeper understanding of the social and economic circumstances of individuals to develop a predisposition for risky behaviors in the context of a zoonotic spillover. Thus, our approach can be used to design better targeted campaigns and can help to prioritize resources in space and time (e.g., vaccination, informative).

Our results reveal the SDE factors most correlated with the infection spillover probability for individuals (Fig. \ref{AIC}).
As expected, the educational and economic levels, the working conditions, and the information access contribute to modulate the risk probability of individuals. Those factors are captured by a reduced number of indicators: work environment, internet use, education background, relative income, water acquisition source, and cellphone ownership. 
Our findings showed that gender, religion, and age do not have a major role in modeling the spillover risk probability. Still, some results are worth mentioning about these demographic indicators.
Young adults (ages between 18-34) and adults (ages between 34-50) constituted $77\%$ of the investigated sample, but they constitute $86\%$  of the respondents at risk.  Also, $50\%$ of the study respondents have an agriculture-related occupation but when computing the percentage within respondents at risk we obtained $79\%$. Thus, our model reveals some small biases that suggest that those age ranges and occupations are more susceptible to risky behaviors related to an Ebola infection spillover. Still, we notice that the size of our sample was relatively small and that a larger sample would be required to show that these biases are significant. Related to this last comment, our methodology leverages efforts made regularly in Sierra Leone to measure the demographics. Ideally, in future survey campaigns additional questions to measure risk predisposition could be included by SSL, similar to those included in our local survey. An increased sample size would allow to refine our results, increase the accuracy, and possibly be analyzed by other quantitative methods that were deemed as inaccurate in our study (i.e., machine learning).

As for how representative is our study to capture the spillover risk probability in rural areas at the national level, the evaluation of the reliability of our data revealed that similar trends were obtained in Bombali and the rest of the country. However, some differences where also observed (Fig.~\ref{demographic}) that might raise questions about the ability to extrapolate our model. 
This is one of the reasons underlying the exploration of different scenarios (Fig.~\ref{map_prob}). 
In that regard, our results are qualitatively robust and show a similar relative risk  among districts.
Nonetheless, we point out that it is certainly possible that if future, larger, surveys are run, other SDE features could be identified as more representative in terms of their predictive capabilities, following the methodology that we propose.
As a possible criticism, the upper and lower bounds of our prediction for the spillover risk probability maps could be considered as too broad: taking as a reference the best fit model,  the resulting probability at a given district is approximately four times larger when the worst case scenario is considered.
Once more, larger data sets would reduce this variability. 

In our study two different factors are integrated when computing the infection spillover exposure map in Sierra Leone: the spillover risk probability and the population density map.
Some districts can actually have a large spillover probability but their low population density helps to diminish their exposure (e.g. Koinadugu). 
The opposite (relative small spillover probability and large population density) can lead to similar spillover exposure levels (e.g. Port Loko). 
Thus, actions should be taken considering the spillover probability as well as the population density of each district.
In any case, our model has identified two districts that because of both individual risk and population density are particularly exposed: Kailahun and Kambia. Taking into account that the 2014 epidemics started in Kailahun, more efforts are still needed to lower the spillover exposure there.

As a matter of discussion, we stress that our study aims at understanding how SDE factors are related with the Ebola spillover risk. However, a more complete picture of the infection spillover map would require to include additional drivers (e.g. ecology effects and bat migration habits). In fact, recent studies have established Ebola spillover risk maps in different regions of the African continent where environmental, climatic, and some anthropogenic factors were considered \cite{lee-cruz_mapping_2021}.
Still, the authors pointed out that there are still important gaps in the knowledge about the factors leading to infection spillover. We believe that our study accounts for some of those factors and envision that the combination of compartmental models able to provide the density of infected animal host driven by enviroclimatic cues \cite{fiorillo_predictive_2018} with our approach would lead to a comprehensive assessment of the risk of spillover.
In this sense, one of the major contributions of this work is the fact that the complete raw data resulting from our survey campaign in Sierra Leone are provided as additional material to this manuscript, for allow other scholars to perform additional analyses.

Effective allocation of resources is necessary to hinder global epidemics, given the limited health care infrastructure in Sierra Leone and other West African nations. 
This requires an established priority of what regions are most at risk and therefore most in need of resources. 
In that regard, our methodology and findings hopefully help to identify the districts which are more susceptible to an infection spillover of Ebola.

\section*{Acknowledgements}
This work is part of the activities of the newly established ``Catastrophe Modeling Center'' at Lehigh University. 
The financial support of the US National Institute of Health through grant 1R15GM123422-01A1 and of Lehigh University through the ``Research Futures: Major Program Development'' grant, through the ``Global Social Impact Fellowship'', and through the ``Mountaintop Initiative'' is gratefully acknowledged.
The assistance of Dr.\ Jessecae Marsh, Prof.\ Khanjan Mehta, Dr.\ Soumyadipta Acharya, and Mr.\ Vaafoulay Kanneh in reviewing the survey instrument is gratefully acknowledged. 
The field and logistical support for the data collection in Sierra Leone provided by World Hope International and by Lehigh's Office of Creative Inquiry was essential to complete the project and is gratefully acknowledged. Particular gratitude goes to the two translators, Mr.\ Salifu Tenneh Samura and Mr.\ Sulaiman Bah.
Finally, the Authors want to acknowledge the support of Lehigh's Interdisciplinary Research Institute for Data, Intelligent Systems, and Computations that through the ``I-DISC Fellow'' program provided the assistance of Mr.\ Dan Luo in the machine learning analysis. J.B. also acknowledges support from the Spanish Ministry of Science and Innovation through grant PID2019-103900GB-I00 and from CSIC through the ``Conexiones-Vida'' Program.
The opinions and conclusions presented in this paper are those of the authors and do not necessarily reflect the views of the sponsoring organizations.

\section*{Author Contributions}

\textbf{Conceptualization:} Sena Mursel, Lindsay Slavit, Nate Alter, Anna Smith, Paolo Bocchini, Javier Buceta. \hfill \break

\textbf{Data curation:} Sena Mursel, Lindsay Slavit, Nate Alter, Anna Smith. \hfill \break

\textbf{Formal analysis:} Sena Mursel, Lindsay Slavit, Nate Alter, Anna Smith, Paolo Bocchini, Javier Buceta. \hfill \break

\textbf{Funding acquisition:} Paolo Bocchini, Javier Buceta. \hfill \break

\textbf{Methodology:} Sena Mursel, Lindsay Slavit, Nate Alter, Anna Smith, Paolo Bocchini, Javier Buceta. \hfill \break

\textbf{Project administration:} Paolo Bocchini, Javier Buceta. \hfill \break

\textbf{Resources:} Sena Mursel, Lindsay Slavit, Nate Alter, Anna Smith, Paolo Bocchini, Javier Buceta. \hfill \break

\textbf{Software:} Sena Mursel, Lindsay Slavit, Nate Alter. \hfill \break

\textbf{Supervision:}  Paolo Bocchini, Javier Buceta. \hfill \break

\textbf{Validation:} Sena Mursel, Lindsay Slavit, Nate Alter, Anna Smith, Paolo Bocchini, Javier Buceta. \hfill \break

\textbf{Writing - original draft:} Sena Mursel, Lindsay Slavit, Nate Alter, Anna Smith, Paolo Bocchini, Javier Buceta. \hfill \break

\textbf{Writing - review \& editing:} Sena Mursel, Lindsay Slavit, Nate Alter, Anna Smith, Paolo Bocchini, Javier Buceta. \hfill \break

\nolinenumbers

%
%
%

\bibliography{bibliography_ebola}



\end{document}